\documentclass[3p, times,merge]{elsarticle}
\journal{Computer Physics Communications}

\usepackage[plain]{algorithm}
\usepackage{amsmath}
\usepackage{amssymb}
\usepackage{bm}
\usepackage[nolabel]{showlabels}
\usepackage[ISO=false]{diffcoeff}
\usepackage{listings}
\usepackage{graphicx}
\usepackage[utf8]{inputenc}
\usepackage{tablefootnote}
\usepackage{threeparttable}
\usepackage{makecell}
\usepackage{xcolor}
\usepackage{textcomp}
\definecolor{mydarkgray}{gray}{0.2}
\definecolor{darkgreen}{rgb}{0,.5,.0}
\usepackage{tikz}

\usepackage{geometry}
\usepackage{subcaption}
\usepackage{hyperref}
\usepackage{caption}
\usepackage{subcaption}
\usepackage{booktabs}

\newcommand{\avg}[1]{\left\langle{#1}\right\rangle}

\lstdefinelanguage{MyPython}{
    keywords={def, for, int, float, return, if, else, range, product},   % ONLY these keywords
    keywordstyle=\color{red}\bfseries,
    comment=[l]{\#},             % Python-style comment
    string=[b]"                 % double-quoted strings
}

\begin{document}

\begin{frontmatter}

\title{Sampling two-dimensional spin systems with transformers}
\author[iis]{Piotr Białas}
\ead{piotr.bialas@uj.edu.pl}
\author[ift]{Piotr Korcyl}
\ead{piotr.korcyl@uj.edu.pl}
\author[ift]{Tomasz Stebel}
\ead{tomasz.stebel@uj.edu.pl}
\author[ift]{Adam Stefański}
\ead{adam.stefanski@student.uj.edu.pl}
\author[iis,ds]{Dawid Zapolski}
\ead{dawid.zapolski@doctoral.uj.edu.pl}

\address[iis]{Institute of Applied Computer Science, Jagiellonian University, ul.~Łojasiewicza 11, 30-348 Kraków, Poland}

\address[ift]{Institute of Theoretical Physics, Jagiellonian University, ul.~Łojasiewicza 11, 30-348 Kraków, Poland}

\address[ds]{Doctoral School of Exact and Natural Sciences, Jagiellonian University, ul.~Łojasiewicza 11, 30-348 Kraków, Poland}

\begin{abstract}
Autoregressive Neural Networks based on dense or convolutional layers have recently been shown to be a viable strategy for generating classical spin systems.
Unlike these methods, sampling with transformers is commonly considered to be computationally inefficient.
In this work, we propose a novel approach to transformer-based neural samplers in which we generate not a single spin per step but groups of spins. As an additional improvement, we construct a model of approximated probabilities, further improving the efficiency of the algorithm. Despite our approach being computationally heavier than dense networks or CNN-based approaches, we were able to sample larger systems of up to $180 \times 180$ spins in case of the Ising model. The Effective Sample Size of our sampler is $\sim 20$ times larger than that of the previous state-of-the-art neural sampler when trained for the $128 \times 128$ Ising model at critical temperature. Finally, we also test our algorithm on the 2D Edwards-Anderson model, where we train $64\times 64$ spin systems.
\end{abstract}


\end{frontmatter}

\section{Introduction}
\label{sect_intr}

  Simulating physical systems is one of the central challenges that led to the development of Monte Carlo (MC) algorithms. Beginning with the pioneering work of Nicholas Metropolis and collaborators~\cite{10.1063/1.1699114}, the past 70 years have witnessed remarkable progress in Markov Chain Monte Carlo (MCMC) methods, resulting in a wide range of applications across science, engineering, finance, and beyond.
Despite this tremendous success, several fundamental limitations of MCMC methods remain. Chief among these are autocorrelations between successive samples and potential issues with ergodicity, which can hinder efficient exploration of complex probability landscapes.
With the advent of modern neural network–based approaches, new opportunities have emerged for the design of more efficient and scalable sampling algorithms, raising hopes of overcoming some of the longstanding challenges associated with traditional MCMC techniques.

Sampling classical spin systems using neural networks was proposed in \cite{2019PhRvL.122h0602W}, where the autoregressive architecture models the conditional probabilities of consecutive spins. The loss function used to train those models is the variational free energy; hence the name of the algorithm: Variational Autoregressive Networks (VAN). In \cite{2019PhRvL.122h0602W} authors used two architectures of autoregressive networks: dense layers with masked connections, known as Masked Autoencoder for Distribution Estimation (MADE) \cite{2015arXiv150203509G} and convolutional layers with masked kernels. The main limitations of the VAN are the unfavorable scaling of numerical cost with the system size and poor training efficiency at large system sizes. For the simplest spin model, the 2-dimensional Ising model, the original VAN cannot be effectively trained for system sizes larger than $32\times 32$ spins. On the other hand, the algorithm is very flexible concerning the type of interactions -- as far as the energy of the spin configuration can be calculated, there are no limitations.

Motivated by the problems of VAN in simulating large system sizes, we proposed in Ref.\cite{Bialas:2022qbs} a modification of this algorithm that utilizes the properties of the nearest neighbors' interactions and the translational symmetry of the Ising model. Hierarchical Autoregressive Networks (HAN) replaced one large dense network from VAN with a hierarchy of much smaller networks, which can be used multiple times to fix the spins in different parts of the configuration. With that change, the algorithm can effectively simulate a 2D Ising system with $64 \times 64$ spins. The HAN algorithm was later applied to the calculations of mutual information \cite{Bialas:2023fjz}, entanglement entropy \cite{Bialas:2024gha}, density matrix \cite{Bialas:2025ldp}, as well as to the simulation of the Potts model \cite{Bialas:2022bdl} and the 3D Ising model \cite{Bialas:2025hxu}. On the other hand, the more complicated models, like spin glasses, cannot be sampled. 

An alternative approach to simulations of spin systems with autoregressive networks by Biazzo et al.~\cite{Biazzo_2023,2024MLS&T...5b5074B} takes into consideration also the form of the interactions between spins. By incorporating parts of these interactions into the model, the authors obtained much better performance of the model compared to VAN (MADE) when simulating spin glass systems. 

Recently, yet another neural network algorithm for sampling the Ising model was proposed \cite{Singha:2025lsd}. It is inspired by the renormalization group technique and performs sampling from low-resolution to high-resolution lattice configurations using conditional probability. Authors report that the Renormalization-informed Generative Critical Sampler (RiGCS) can effectively sample 2D Ising systems of size $128\times 128$ spins. 

The above-mentioned algorithms utilize the physical properties of spin systems, but the neural network architectures that are used are usually just dense or convolutional networks. In this contribution, we shall present a version of the VAN algorithm that is based on the transformer architecture \cite{NIPS2017_3f5ee243}. We call our algorithm "transformer VAN" (tVAN), to distinguish it from the original proposal. We show that even with a relatively small transformer model, a sampler can efficiently simulate the 2D Ising model with $180\times 180$ spins and the spin glass system with $64 \times 64$ spins. With such system sizes trained, tVAN is a state of the art neural sampler of spin systems. It enjoys the flexibility of the original VAN concerning the variety of interactions it can simulate.

We test our tVAN architecture on the 2D classical spin system with a square lattice of size $L\times L$ spins and periodic boundary conditions. The spins have values $s^i=\pm 1$. The energy of the configuration is given by:
\begin{equation}
    E(\mathbf{s}) = -  \sum_{\langle i,j \rangle} J_{i,j} s^i \, s^j \,,
\label{spin_hamilt}
\end{equation}
where the sum is performed over the nearest neighbor interactions. Our default choice in this manuscript is the Ising model, where $J_{i,j}=1$ for all $i,j$. We shall also consider the Edwards-Anderson (EA) spin glass model with $J_{i,j}=\pm 1$, where the value of each $J_{i,j}$ is sampled with probability $1/2$. The probability of the spin configuration is given by the Boltzmann distribution:
\begin{align}
    p(\mathbf{s}) = \frac{1}{Z} \exp(-\beta E(\mathbf{s})) \,,
    \label{eq_boltz_distr}
\end{align}
where $\beta$ is an inverse temperature, $\beta=1/k_BT$; $Z$ is the partition function, $Z=\sum_{\mathbf{s}} \exp(-\beta E(\mathbf{s}))$, and the sum is performed over all $2^{L^2}$ configurations. The free energy is defined as:
\begin{equation}
F= -\frac{1}{\beta} \ln Z.
\label{F_def_eq}
\end{equation}

In the Ising model, the exact value of $F$ can be calculated  for arbitrary $\beta$ and $L$ \cite{PhysRev.185.832}. We take it as our benchmark, which we compare with variational values obtained from the models, $F_q$. The Ising model in 2D exhibits a second order phase transition when $L\to\infty$. The critical inverse temperature is given by $\beta_c=\frac{1}{2}\ln(1+\sqrt{2})$. In what follows, when considering the Ising model, we use $\beta=\beta_c$.

In case of the spin glass models like the EA model considered here, one has to average over the link variables $J_{ij}$ and equation \eqref{F_def_eq} becomes
\begin{equation}
\label{eq:F-J}
F= -\frac{1}{\beta} \avg{\ln Z}_{J}.
\end{equation}
As the aim of this reference is not the study of spin glasses but a proof of concept of the method, we only present results for several sets of coupling $J$ without any averaging. In general, the spin glass models do not exhibit spontaneous symmetry breaking but rather a glassy transition associated with replica symmetry breaking \cite{Edwards_1975,Parisi1983}. However, the two-dimensional system considered here only exhibits a phase transition at $T=0$ ($\beta\rightarrow\infty$) \cite{PhysRevB.22.288} and is increasingly harder to simulate with increasing $\beta$.

\section{Neural samplers}
\label{sect_NS}

The key idea behind the neural samplers is that a neural network algorithm can generate states of a statistical (or lattice field theory) system. The algorithm samples states from a probability distribution $q_{\theta}$ (where $\theta$ denotes a set of parameters of the neural network) that should be as close as possible to the target probability $p$. The architecture is expected to provide explicit probabilities of the generated samples\footnote{Note that some generative architectures, e.g.~Generative Adversarial Networks (GANs) \cite{2014arXiv1406.2661G}, are ruled out by this requirement.}. This is important because, in general, neural algorithms are not able to perfectly match the probability distribution $p$, and the difference between $p$ and $q_\theta$ leads to a bias for observables. If the architecture provides the probability of the generated sample, one can apply algorithms correcting the difference between $p$ and $q_\theta$; these are known as Neural Markov Chain Monte Carlo \cite{PhysRevD.100.034515} or Neural Importance Sampling \cite{2020PhRvE.101b3304N}.

The training of neural samplers is usually performed in the so-called reverse mode, when no external data are necessary. The difference between the two distributions $q_\theta$ and $p$ is quantified by the (reverse) Kullback-Leibler divergence:
\begin{equation}\label{eq-KL}
    D_{KL}(q_{\theta}|p) = \sum_{\mathbf{s}} \, q_{\theta}(\mathbf{s}) \ln \frac{q_{\theta}(\mathbf{s})}{p(\mathbf{s})} = \avg{\ln q_{\theta}(\mathbf{s})-\ln p(\mathbf{s})}_{q_{\theta}}.
\end{equation}
One can show that $D_{KL}(q_{\theta}|p)\ge0$ and $D_{KL}(q_{\theta}|p)=0 \Leftrightarrow q_{\theta}=p$. Substituting Eq.~(\ref{eq_boltz_distr}) into (\ref{eq-KL}) one gets $D_{KL}(q_{\theta}|p)=\beta(F_q-F)$ where $F$ is the free energy Eq.~(\ref{F_def_eq}) and
\begin{equation}
    F_q=\frac{1}{\beta}\avg{\ln q_{\theta}(\mathbf{s})+\beta  E(\mathbf{s}) }_{q_{\theta}}
    \label{loss_function}
\end{equation}
is the variational free energy satisfying the condition $F_q\ge F$. Since $F$ is a constant, minimizing $F_q$ is equivalent to minimizing $D_{KL}$; however, to calculate $F_q$ one does not need the value of $F$ which is often difficult to obtain. In practice, the average $\avg{\ldots }_{q_{\theta}}$ is performed with a relatively small batch of samples\footnote{In this paper, we shall use $N_{batch}\sim10^3-10^4$ for training and  $N_{batch}=10^6$ for the final evaluation of a given quantity.} generated from the distribution $q_{\theta}$:
\begin{equation}
    \avg{\ldots }_{q_{\theta}} \to \frac{1}{N_{batch}}\sum_{i=1}^{N_{batch}} (\ldots), \ \textbf{s}_{i}\sim q_\theta.
    \label{averaging_over_q}
\end{equation}
The training requires calculating the gradient of the loss function (\ref{loss_function}) with respect to the network weights $\theta$. In this manuscript, we shall follow the prescription proposed in Ref.~\cite{2019PhRvL.122h0602W}:
\begin{equation}
    \frac{d F_q}{d\theta}= \frac{1}{\beta}\avg{\left(S-\avg{S}_{q_{\theta}} \right) \frac{d \ln q_\theta}{d\theta} }_{q_{\theta}}, 
\end{equation}
where $S=\ln q_{\theta}(\mathbf{s})+\beta  E(\mathbf{s})$.
The update of weights is called an epoch. Typically, training consists of $10^4-10^5$ epochs.

There are many ways to assess the quality of training. Here, we should use two that do not rely on external data. The first is 
variational free energy (\ref{loss_function}) per spin - the smaller $F_q/L^2$, the closer $q_\theta$ is to $p$. As we mentioned before, in the case of the Ising model, $F$ can be calculated analytically, which allows us to compare $F_q$ with the exact value of free energy. The second measure, which is often used for the evaluation of neural samplers' training, is the Effective Sample Size (ESS) \cite{Liu, Kong}:
\begin{equation}
    \text{ESS}= \frac{\avg{\hat{w}(\mathbf{s}_i)}_{q_\theta}^2}{\avg{\hat{w}(\mathbf{s}_i)^2}_{q_\theta}},\  \textrm{ where } \ \hat w(\mathbf{s}_i) = \frac{e^{-\beta E(\mathbf{s}_i)}}{q_\theta(\mathbf{s}_{i})}
\label{ESS_definition}
\end{equation}
and the average over $q_\theta$ is performed as explained in (\ref{averaging_over_q}). It is easy to see that $0<\text{ESS}\le1$ and $p=q_\theta \Rightarrow \text{ESS}=1$.

\subsection{Autoregressive networks in statistical systems}

In the context of statistical spin systems, the most popular samplers are based on autoregressive architectures, where one expresses the $q_\theta$ as a product of conditional probabilities of consecutive variables. In the original VAN algorithm \cite{2019PhRvL.122h0602W}, the probability density is factorized into conditional probabilities of single spins:
\begin{equation}\label{eq:cond_probs}
    q_\theta(\mathbf{s}) = q_\theta(s^1)\prod_{i=2}^{L^2} q_\theta(s^i| s^{<i}),
\end{equation}
where $s^{<i}=(s^1,s^2,\ldots., s^{i-1})$.
The configuration is generated by ancestor sampling, where the network is evaluated $L^2$ times. During the $i$-th evaluation, the conditional probability of the $i$-th spin, $q_\theta(s^i| s^{<i})$, is calculated based on the knowledge of all $i-1$ spins that have already been fixed. Having $q_\theta(s^i| s^{<i})$, the $i$-th spin is drawn.

\subsection{Transformer VAN (tVAN)}
\label{tVAN_descript}

\begin{figure}
    \centering
        \includegraphics[width=\linewidth]{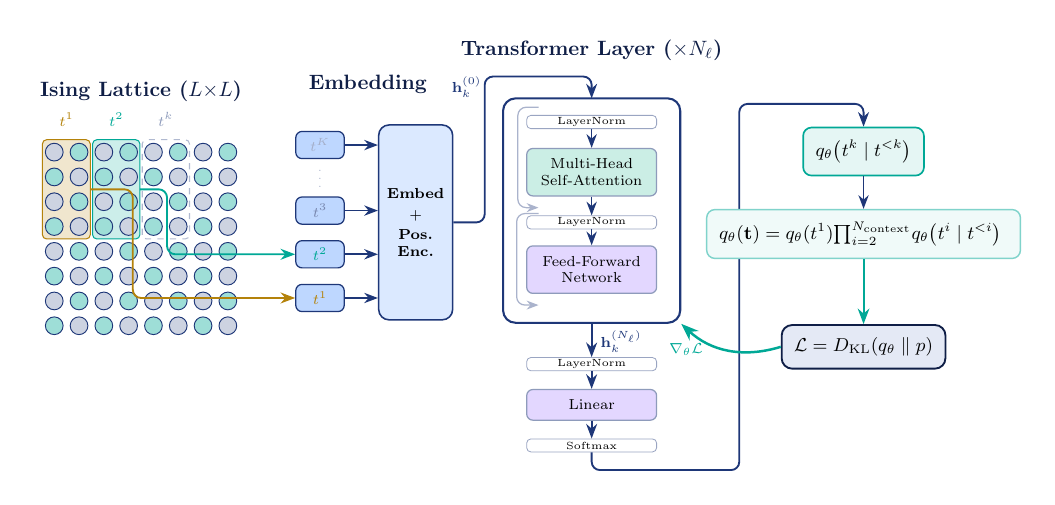}
    \caption{Schematic view of tVAN algorithm. Configuration of spins is divided into patches that are generated sequentially. Output of the transformer is a conditional probability of the next patch.}
        \label{fig:tVAN_view}
\end{figure}

The Transformer architecture \cite{NIPS2017_3f5ee243} is widely used in Deep Learning, where it has revolutionized Natural Language Processing (NLP). The text is converted into a series of tokens that are processed by the transformer. The number of possible tokens is called the vocabulary size, and we shall denote it with $N_{\rm vocab}$. 
In decoder-only mode, which we will consider in this manuscript, at each iteration, a new token is proposed based on the values of the previously generated tokens. To be more specific, the output of the transformer in the $i$-th iteration is a list of probabilities of $N_{\rm vocab}$ possible tokens. As such, this architecture is exactly the autoregressive network described above, so applying it to spin generation is straightforward. Since configurations always have a fixed size, the series of tokens representing it has a fixed length, which we denote $N_{\rm context}$. In NLP, the context length can be different from the length of the token series; however, here we keep them equal.

\begin{figure}
    \centering
\begin{tikzpicture}[scale=0.8]
  % Define parameters
  \def\circleradius{0.3}
  \def\spacing{1}

  \pgfmathtruncatemacro{\num}{0}
  \xdef\num{\num}
  % Draw circles with numbers
  \foreach \row in {0,...,3} {
    \foreach \col in {0,1} {
      \foreach \srow in {0,1} {
        \foreach \scol in {0,...,3} {
          \pgfmathtruncatemacro{\nrow}{\row * 2 + \srow}
          \pgfmathtruncatemacro{\ncol}{\col * 4 + \scol}
          
          \ifthenelse{\num > 31}{\node[draw, circle, minimum size=2*\circleradius*10mm,
                inner sep=0pt, fill=white, name=node\num, dashed] at (\ncol*\spacing, -\nrow*\spacing) {\small \num};}{
                \ifthenelse{\num > 23}{
                \node[draw, double, circle, minimum size=2*\circleradius*10mm,
                inner sep=0pt, fill=white, name=node\num] at (\ncol*\spacing, -\nrow*\spacing) {\small \num};
                }{
                          \node[draw, circle, minimum size=2*\circleradius*10mm,
                inner sep=0pt, fill=white, name=node\num] at (\ncol*\spacing, -\nrow*\spacing) {\small \num};
                }
                }
          \pgfmathtruncatemacro{\tmpnum}{\num + 1}
          \xdef\num{\tmpnum}
        }
      }
    }
  }
  
  % Draw rectangles (2 rows x 4 columns of rectangles)
  % Each rectangle contains 4x2 circles
  \def\xlen{4}
  \def\ylen{2}
  \foreach \rectrow in {0,1,2,3} {
    \foreach \rectcol in {0,1} {
      % Calculate rectangle corners
      \pgfmathsetmacro{\xmin}{(\rectcol*\xlen)*\spacing - 0.5*\spacing + 0.05}
      \pgfmathsetmacro{\xmax}{(\rectcol*\xlen + \xlen)*\spacing - 0.5*\spacing - 0.05}
      \pgfmathsetmacro{\ymin}{-(\rectrow*\ylen)*\spacing + 0.5*\spacing - 0.05}
      \pgfmathsetmacro{\ymax}{-(\rectrow*\ylen + \ylen)*\spacing + 0.5*\spacing + 0.05}
      % Draw rectangle
      \draw[gray, rounded corners=2mm] (\xmin, \ymin) rectangle (\xmax, \ymax);
    }
  }
\draw[red, thick]   (node12)  -- (node24);
\draw[red, thick]   (node19)  -- (node24);
\draw[red, thick]   (node23)  -- (node28);
\draw[red, thick]   (node13)  -- (node25);
\draw[red, thick]   (node14)  -- (node26);
\draw[red, thick]   (node15)  -- (node27);

\foreach \row in {24,28}{
    \foreach \col in {0,...,2}{
      \pgfmathtruncatemacro{\num}{\row + \col}
      \pgfmathtruncatemacro{\numinc}{\num + 1}
      \draw[green, thick] (node\num) -- (node\numinc);
    }
}
\foreach \col in {24,...,27} {
    \pgfmathtruncatemacro{\num}{\col + 4}
    \draw[green, thick] (node\col) -- (node\num);
}
\end{tikzpicture}
    \caption{The sample is generated sequentially, starting from the upper-left corner. Grey rectangles denote the patches produced by the transformer in a single step. Assuming that three initial patches have already been generated, probability of fourth patch is calculated. Green connections indicate interactions between spins inside a patch, while red represents interactions with neighboring patches; sum of those is $E^4$, Eq.~\ref{approx_energies}, and enters approximate probabilities formula, Eq.~\ref{ap_equation}.}
    \label{fig:patches}
\end{figure}
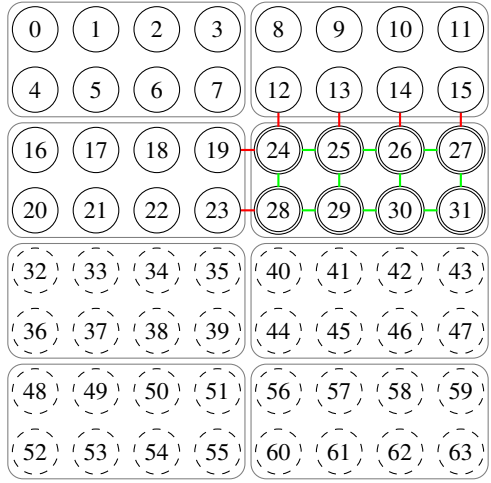

 The idea of tVAN is presented in Fig.\ref{fig:tVAN_view}. The simplest approach is to treat each  spin as a single token, hence $N_{\rm vocab}=2$ and $N_{\rm context}=L^2$. At each evaluation of the transformer, one spin is fixed, similarly to the original VAN algorithm. Such architecture is computationally expensive unless $L$ is small, due to the large context length. An alternative approach, inspired by vision transformers \cite{dosovitskiy2020image}, is to group $r \times c$ spins together into patches $\bm{t}=(t^1,t^2,\ldots,t^{N_{\rm context}})$, where $N_{\rm context}=L^2/(r\times c)$. Then we treat each patch $t^i$ as a single token with $N_{\rm vocab}=2^{r\times c}$. Eq.~\ref{eq:cond_probs} can be written in terms of patches of spins:
\begin{equation}
    q_\theta(\mathbf{t}) = q_\theta(t^1)\prod_{i=2}^{N_{\rm context}} q_\theta(t^i| t^{<i}), 
\end{equation}
This idea was also presented in Fig.~\ref{fig:patches}, where we show an example of a system with $8\times 8$ spins that is divided into patches of sizes $2\times 4$. 
As the vocabulary size grows exponentially with patch size, it is clear that patches need to be rather small compared to $L^2$ (see \ref{scaling_derivation_details} for more details). However, it is hard to predict  the optimal size of the patch for a given $L$; it needs to be determined by numerical experiments. 

In decoder-only mode, one needs to generate a first token $t^1$ before the first evaluation of the transformer. This means that the probability distribution of the first token, $q_\theta(t^1)$, needs to be modeled differently than the transformers output. We use the parametrization where the probability that $t^1$ takes the $j$-th value is given by:
\begin{equation}
    q_\theta(t^{1}_j) = \text{softmax}_j(x_j),
\end{equation}
where 
$x_1,x_2,\ldots x_{N_{\rm vocab}}$ are learnable parameters. In the simplest case, they are initialized by drawing from a uniform distribution. The softmax function guaranties that $q(t^{1}_j)>0$ for all $j$ and $\sum_{j=1}^{N_{\rm vocab}} q(t^{1}_j) =1$. 
Subsequent tokens are generated iteratively based on the outputs of the transformer $f$:
\begin{equation}
q_\theta(t^{i}_{j}|t^{<i}) = \text{softmax}_{j}\left( f_{j}(t^{<i}) \right).
\label{no_AP_con_prob}
\end{equation}

At the architectural level, the transformer consists of multiple blocks connected sequentially, each containing a self-attention mechanism and a feed-forward neural network(see Figure \ref{fig:tVAN_view}). The self-attention mechanism introduces correlation between embedded tokens, while the feed-forward network changes the representations of the tokens independently. The embedding is a learnable map between tokens and $n_{\rm embed}$-dimensional vectors. Our implementation of the transformer block is based on the nanoGPT model \cite{nanoGPT}, and we have added a KV cache to reduce generation time. Details of the architecture are given in \ref{impl_details}.

\subsection{Approximate probabilities (AP)}
\label{AP_section}

To improve learning speed, one can make use of some physical properties of  the system. In this manuscript, we shall use the simplest approximation of probabilities (see e.g.~\cite{Biazzo_2023} where similar ideas were applied), which takes into account the internal energy of the patch spins -- see the green connections in Fig.~\ref{fig:patches} -- and the energy of interactions between the currently generated patch and its two nearest neighboring patches that are already fixed (on the left and above) -- red connections in Fig.~\ref{fig:patches}. Hence, we define for the $i$-th patch: 
\begin{equation}
E^i= - \sum_{{\rm internal}} J_{kj} s^k s^j - \sum_{{\rm neigh. \, of} \ i} J_{kj} s^k s^j,    
\label{approx_energies}
\end{equation}
where the sums are performed over internal connections and the connections with the nearest neighboring spins of the patch. For the first patch, we sum only internal spin connections. When the patch is a single spin, $r=c=1$, only connections from the neighbors are summed. Such approximation may be poor as stand alone; however, one can use it to accompany the transformer output so that the neural network needs to fill the gap between the approximation and the true value. 

With AP we get the following conditional probabilities:
\begin{equation}
q(t^{i}_{j}|t^{<i}) = \text{softmax}_{j}\left(-\beta E^i(t_j) + f_{j}(t^{<i})\right).
\label{ap_equation}
\end{equation}
where for each possible token $j\in \{1, \ldots,N_{\rm vocab} \}$, we need to calculate the energy (\ref{approx_energies}), which we denote as $E^i(t_j)$. The first token, which is not generated by the transformer, is drawn from the probability: 
\begin{equation}
      q(t^{1}_j) = \text{softmax}_j(-\beta E^1(t_j) + x_j),  
 \label{ap_equation_first}     
\end{equation}
where $x_1,x_2,\ldots x_{N_{\rm vocab}}$ are learnable parameters, as before.

\section{Numerical results for Ising model}
\label{sec: results}

\subsection{Patch shape}
\label{prob_effect}

\begin{figure}
    \centering
    \includegraphics[width=1\linewidth]{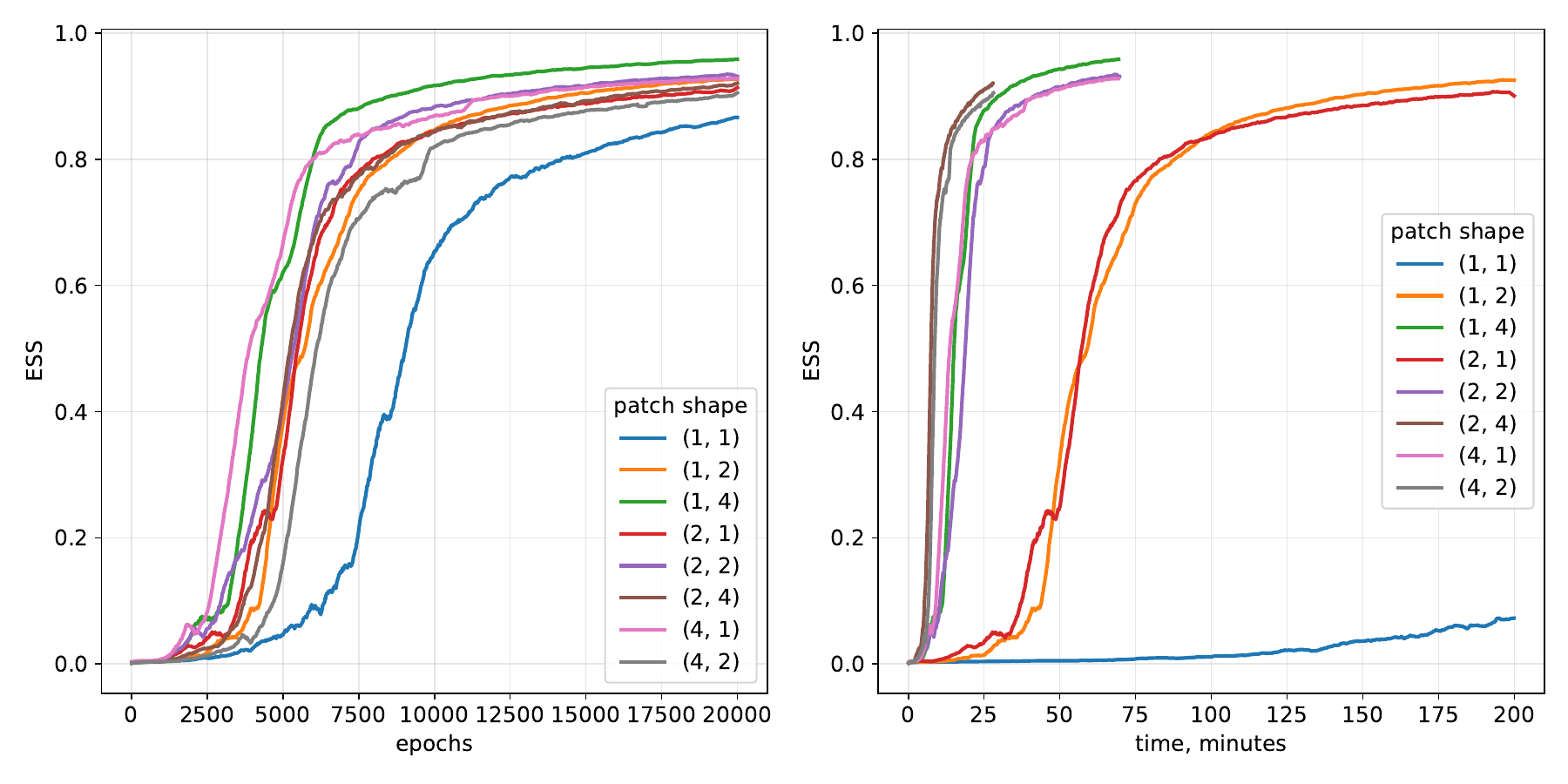}
    \caption{The history of Ising model training for $L=32$ with different patch shapes. No approximate probabilities were used. The lines show moving average of ESS (with window of 30 points, points were saved every 10 epochs) that are averaged over 3 or 4 independent runs. Left: ESS as function of epoch. Right: the same, but as function of runtime.}
    \label{fig:ess_patches_32}
\end{figure}

We first examine the performance of our algorithm for different patch sizes. In Fig.~\ref{fig:ess_patches_32} we show ESS as a function of epoch (left panel) and time (right panel) for several shapes of the patch for system size $L=32$. No AP were used. We see that the least effective method is single spin generation (patch $1\times1$): the total time to train the model to $\text{ESS}>0.8$ is much larger than for other patch sizes. This is mostly due to the slowness of the algorithm (time per epoch). Effectiveness grows with patch size, and the fastest training (in wall time) is for shapes $2\times 4$ and $4 \times 2$. As we move to larger patch sizes ($4\times 4$), the training time drastically rises (not shown in the figure), as the time per epoch is very large. We also notice that the orientation of the patch (shape $2\times 4$ vs $4\times 2$) does not influence training efficiency.\footnote{We noticed that the orientation of the patch has some impact when using AP, patches with $r<c$ seem to lead to better training.}

In Fig.~\ref{fig:ESS_F_L=120}, we see that for system $L=120$, patch $3\times4$ is more effective than $2\times4$. All runs took 48 hours and depending on the speed of the algorithm, the number of epochs in the run could vary. %The learning rate was set to 0.0003 and batch size to 4096.
For this system size, the algorithm is $\sim4$ times faster for the patch size $3\times4$ than for $2\times4$. For patch $3\times5$, the vocabulary size is 32,768, which makes the model significantly larger in terms of the number of parameters and memory requirements (we are not presenting results for this case). 

We conclude that for a system size of order $L\sim100$ and smaller, the training is most effective for patches with $8-12$ spins\footnote{When the patch is of rectangular shape, $L$ needs to be divisible by the patch dimensions; thus, some patch sizes are excluded.}.

\subsection{Approximate probabilities}

\begin{figure}
    \includegraphics[width=\textwidth]{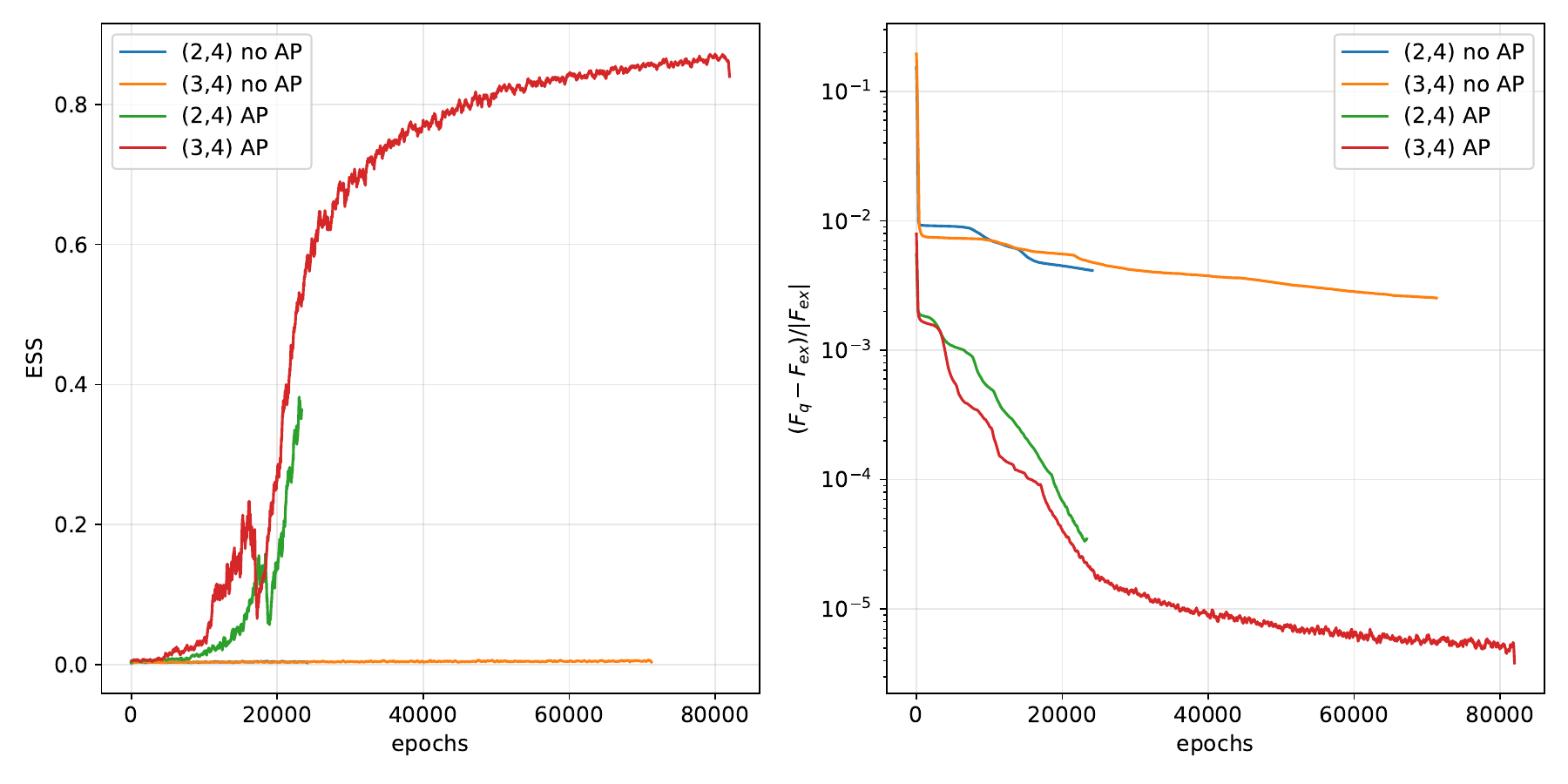}
    \caption{The history of Ising model training for system size $L=120$, for two patch shapes,  $2\times4$ and $3\times4$, with and without approximate probabilities. Moving average is performed over 30 consecutive points and points were saved every 10 epochs. All runs took 48 hours. Left: ESS as function of epoch. Right: Relative values of $F_q/L^2$ compared to $F_{ex}/L^2=-2.10975198.$}
    \label{fig:ESS_F_L=120}
\end{figure}

 In Fig.~\ref{fig:ESS_F_L=120} we also show the effect of adding approximate probabilities (AP). In the left panel, we present ESS as a function of training epochs. To reduce statistical fluctuations, we present the moving averages of ESS history with a window size of 30 points. As is clear from the plots, within 48 hours, transformers without AP did not reach an ESS significantly larger than 0. On the other hand, with AP, the transformer learns much more efficiently, reaching ESS$>0.8$. 
Additionally, AP does not significantly slow down the algorithm. Relative values of $F_q$ compared to $F_{ex}$ are shown in the right panel of Fig.~\ref{fig:ESS_F_L=120}. As expected, $F_q$ decreases much faster when AP is used, reaching a $O(10^{-5})$ difference from the exact value. One should note that even w/o AP, the transformer is improving during training. As we will see in the next subsection, it is possible to get $\text{ESS}>0.5$ w/o AP, but it requires more training time.
 
\subsection{Comparison of autoregressive architectures}
\label{comparison_with_other_arch}

In this section, we show the performance of the tVAN architecture regarding its ability to learn the Boltzmann distribution of the Ising model. We compare it with the HAN \cite{Bialas:2022qbs} and RIGCS \cite{Singha:2025lsd} algorithms. 

We start with system size $L=128$ as it is the largest size that was simulated using all three architectures. In Table \ref{tab:van_han_rigcs}, we present the final values of training quality measures. We compare the ESS values and the relative difference between $F_q$ and the exact value for $\beta=\beta_c$, which is $F_{ex}/L^2=-2.10973977$. 
Both $\text{ESS}$ and $F_q$ were measured using $10^6$ configurations generated after training was finished. With this number of samples, the statistical errors of the measurements are negligible.

We note first that the HAN fails to train on such a system size. The training time is shorter than for tVAN, as HAN algorithm is relatively fast (time per epoch). On the other hand, the loss function for HAN does not improve when increasing the training time. The ESS is very low and cannot be precisely determined: the variance of weights in this case is divergent; hence, when measured on the finite batch, ESS will strongly depend on its size. Such behavior is often encountered when the supports of $q_\theta$ and $p$ do not match. For the relative difference between $F_q$ and $F$, we obtain a result of the order of $10^{-4}$, which may seem small; however, for $L=128$, this translates to a factor of 10 difference between the estimated and true value of the partition function $Z$. 

The RiGCS architecture performs better than HAN; it reproduces the support of $p$ (see the discussion in the original paper \cite{Singha:2025lsd}). On the other hand, the final ESS is 0.03, which affects the effective statistics of data generated with this method. One should note that \cite{Singha:2025lsd} was using $\beta=0.44$ which is slightly smaller than our $\beta_c=0.44068679\ldots$

Now, we present the performance of three different architectures of tVAN: tVAN with one transformer layer (our default option in this manuscript), tVAN with two transformer layers, and tVAN with AP (one transformer layer). The patch size is $2\times4$ and the batch size is $8192$. As expected, the best final measures were obtained when AP were used; ESS reaches a value of 0.84. Even without using AP, we were able to train tVAN to $\text{ESS}>0.4$. We also note that adding a second layer of the transformer significantly improved the performance of the model ($\text{ESS}\approx0.7$).

\begin{table}
    \centering
    \begin{tabular}{lcccc}
        \toprule
        algorithm & system size $L$ & time of training  & $(F_q-F)/|F|$ & ESS \\
        \midrule
        HAN& 128 & 48h & $1.5 \times 10^{-4}$  & $<10^{-3}$ \\
        RiGCS ($\beta=0.44$) & 128 & 14.5h (single A100 GPU) & $1.1 \times 10^{-4}$ & $0.03$  \\  
        tVAN one layer & 128 & 228h & $2.5 \times 10^{-5}$ & 0.43  \\
        tVAN two layers & 128 & 267h  & $1.1 \times 10^{-5}$  & 0.68 \\
        tVAN one layer + AP & 128 & 82h  & $5.5 \times 10^{-6}$ & 0.84 \\
        tVAN one layer + AP & 180 &  192h & $8.8\times 10^{-6}$ &  0.59  \\ 
        \bottomrule
    \end{tabular}
    \caption{Results for training for Ising model using different algorithms at $\beta_c\approx 0.44068679$. RiGCS results are taken from Ref.~\cite{Singha:2025lsd} where $\beta=0.44$ was used and training was performed on different hardware.  }
    \label{tab:van_han_rigcs}
\end{table}

\begin{figure}
    \centering
    \includegraphics[width=1\linewidth]{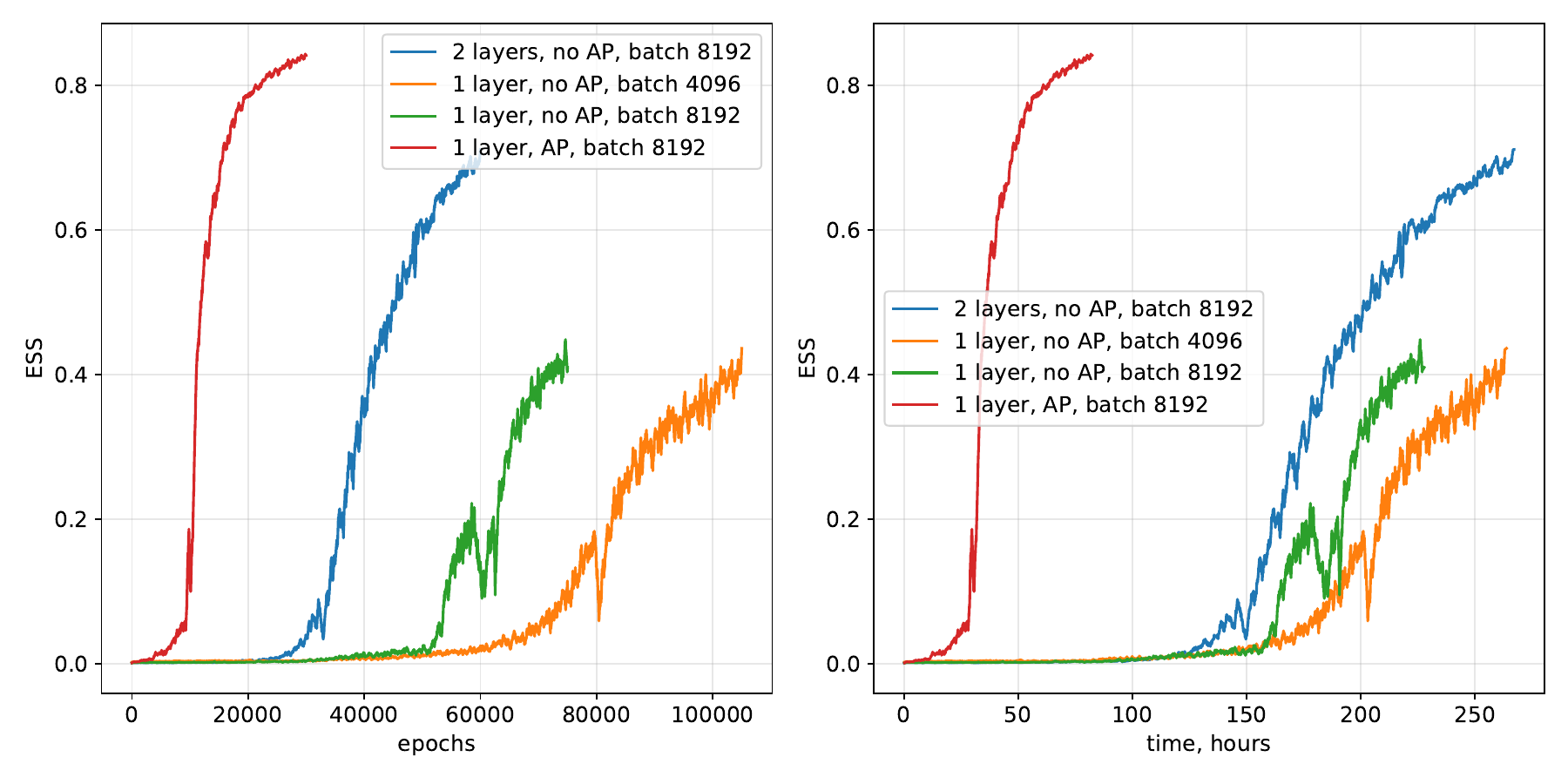}
    \caption{Comparison of ESS during training using different setups. Results are for Ising model with $L = 128$. Moving average is performed over 30 consecutive points and points were saved every 10 epochs.}
    \label{fig:comp}
\end{figure}

In Fig.~\ref{fig:comp} we show the history of training of the tVAN architectures: $\text{ESS}$ as a function of the number of epochs (left panel) and wall time (right panel). We see a clear advantage to using AP. We also note that using two layers of the transformer leads to higher $\text{ESS}$ in the same wall time, even though the algorithm is clearly slower (time per epoch). Finally, we show the run of tVAN (no AP, one layer) with a smaller batch size, $N_{batch}=4096$. Clearly, the larger batch is beneficial. Looking at the history of $\text{ESS}$ during training, we see that all architectures were still improving at the time the training was stopped. Hence, the final values of performance measures, as presented in Table \ref{tab:van_han_rigcs}, could be better; however, this comes at the cost of additional training time. It is worth noting that for $L=128$ one cannot use a patch size $3\times4$ that is more efficient for system sizes of this order: as we showed in Fig. \ref{fig:ess_patches_32}, for $L=120$ one reaches $\text{ESS}\approx 0.7$ after 20h of training (when AP are used).

In the last row of Table \ref{tab:van_han_rigcs}, we present training performance measures obtained for $L=180$. This is roughly twice bigger system (concerning the number of spins) than $L=128$ and, to our knowledge, such a large system size has not been trained before in the literature. We used AP, one transformer layer and patch size $3 \times 4$, which is optimal in this case. tVAN reached $\text{ESS}=0.59$ and $(F_q-F)/|F|=8.8\times 10^{-6}$ with 8 days of training. 

\section{Numerical results for spin glass}
\label{spin_glass}

 Now we will give the reader a glimpse of how the EA model differs from the Ising model regarding the training of the neural sampler. A detailed presentation of tVAN performance in training spin glass distributions will be presented elsewhere. First, we note that in this manuscript we restrict ourselves to the training of the Boltzmann distribution at several sets of couplings $J_{i,j}$. Namely, a set of couplings $J$ is drawn before the training, fixing some target Boltzmann distribution $p_J$, and it does not change throughout the entire training procedure. Therefore, $q_\theta$ has a fixed target probability distribution $p_J$.   

The tVAN (or VAN in general) algorithm can be applied to any type of interaction described by energy $E(\textbf{s})$. The only possible restriction is the application of approximate probabilities described in section \ref{AP_section}: when the interactions are beyond nearest neighbors (long range), the approximation we used may not be useful. In this case, one can use tVAN without AP or develop some other physically-motivated approximation. In this section, we focus on the EA model, where only nearest neighbor interactions are allowed; hence, the AP are expected to work, and we applied them here. We focus on the $L=64$ system, as the EA model is harder to train than the Ising model.

The architecture of tVAN applied to the EA model is almost the same as that for the Ising model. The only complication comes from the fact that the approximate probability of a given patch depends not only on the spins but also on the couplings $J$. 

\begin{figure}
\begin{subfigure}[b]{0.5\textwidth}
    \includegraphics[width=\textwidth]{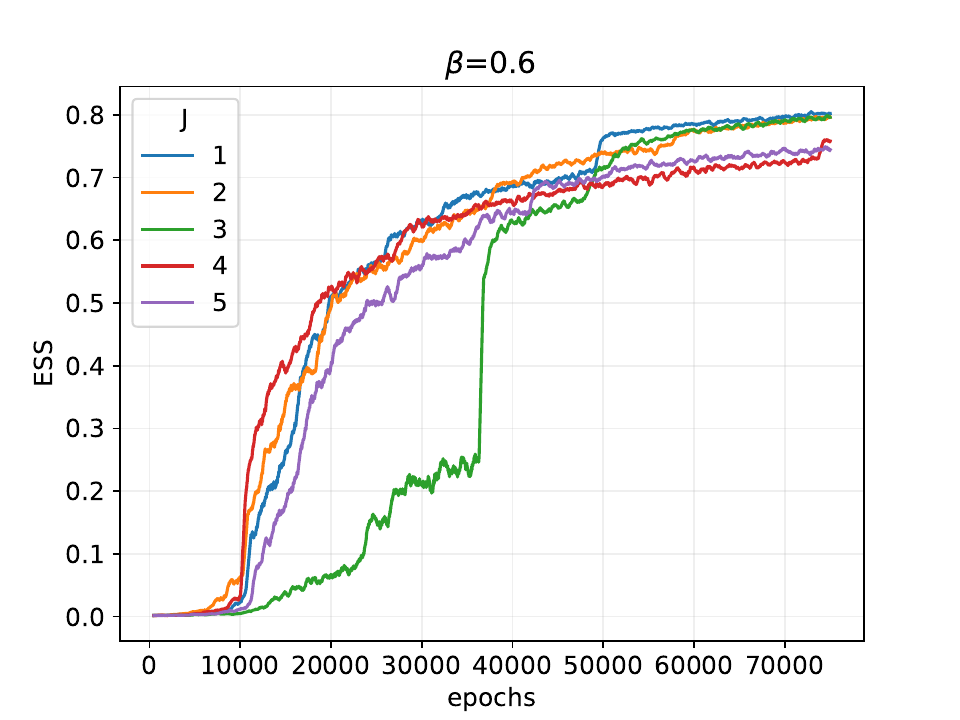}
\end{subfigure}
\begin{subfigure}[b]{0.5\textwidth}
    \includegraphics[width=\textwidth]{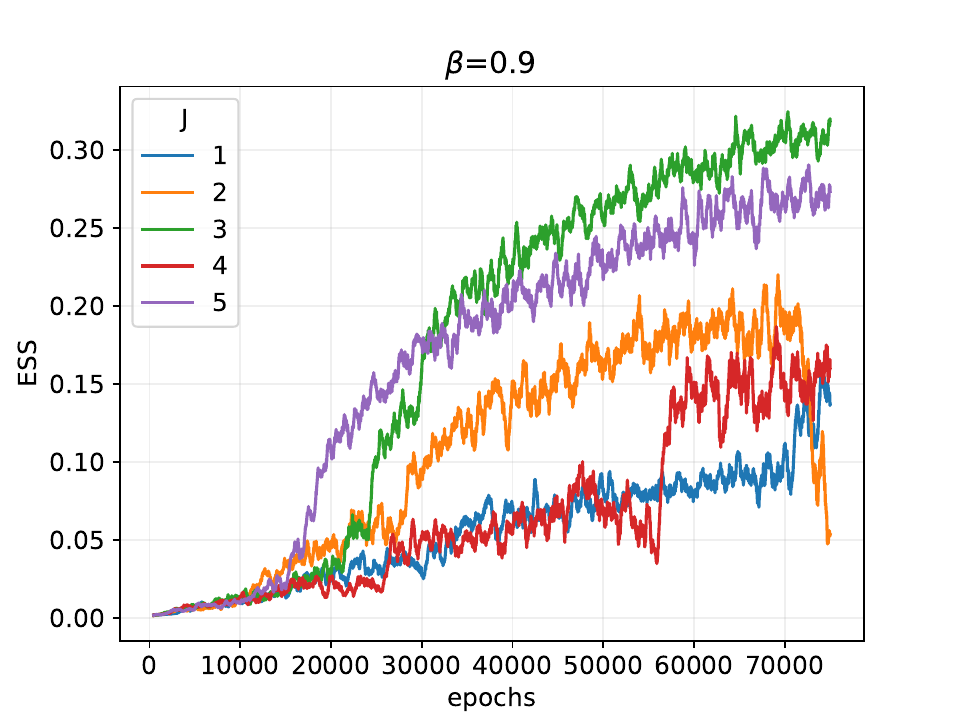}
\end{subfigure}
    \caption{The history of Edwards-Anderson model training for system size $L=64$. We show $\text{ESS}$ as function of epoch for 5 different sets of couplings $J$ (curves with different colors). All the runs took 48h. Training was performed for two inverse temperatures: $\beta=0.6$ (left panel) and $\beta=0.9$ (right panel). Moving average is performed over 50 consecutive points and points were saved every 10 epochs.}
    \label{fig_ESS_spin_glass_0609}
\end{figure}

In Fig.~\ref{fig_ESS_spin_glass_0609}, we show the training history of $\text{ESS}$ for two values of inverse temperature, $\beta=0.6$ (left panel) and $\beta=0.9$ (right panel).\footnote{We also tried smaller values of inverse temperatures, $\beta=0.1$ or $0.3$ and found worse performance of tVAN with AP than without them when considering $L=64$ (for $L\le32$ AP seem to help in training). This unexpected observation requires further investigation.} We drew 5 different sets of coupling $J$ (distinguished by colors in Fig.~\ref{fig_ESS_spin_glass_0609}) and for each set, we performed training using two temperatures. Each run took 48 hours, and all the hyperparameters were kept the same. 

Comparing results for both temperatures, we notice that at larger $\beta$ the system is harder to train. For $\beta=0.6$, we reach $\text{ESS}$ between 0.7 and 0.8, whereas for $\beta=0.9$, $\text{ESS}$ does not exceed 0.3. Further training could improve efficiency, but we leave it for future studies. We notice some differences between $J$ sets, but this is not a large variation and could be a result of statistical fluctuations experienced during training. As there are no analytical results for this model, we compared the tVAN distribution of samples with Monte-Carlo simulations (performed using the parallel tempering technique) and found that the supports of $q_\theta$ and $p$ are equal.

\section{Summary}
\label{summary}

In this manuscript, we used a transformer to sample 2d classical spin systems. For this purpose, we used the idea of Variational Autoregressive Networks \cite{2019PhRvL.122h0602W} but replaced the original dense/convolutional networks with a transformer architecture. The crucial step to obtain good performance was grouping spins into patches (of 8 to 12 spins), that can be generated as one token by the transformer. This significantly speeds up the algorithm and improves its ability to train on large system sizes. We also introduced approximate probabilities, which work for nearest neighbor interactions and lead to better performance of our models.

We have checked that for Ising systems with $128\times 128$ spins, our tVAN algorithm outperforms other state-of-the-art neural samplers. For the first time, we trained a neural sampler for a system with sizes $180\times 180$. For the Edwards-Anderson model, we were able to achieve good performance of the sampler with $64\times 64$ spins, although this was done for a fixed instance of $J$ couplings and a relatively small inverse temperature $\beta$. The results of this manuscript prove that the transformer can be a good sampler of the physical system, which somewhat contradicts the opposing conclusions from previous works \cite{ Abbott:2023thq,PhysRevE.111.025304,Singha:2025lsd}.

There are several directions to extend this work. First, one can explore more complicated physical models. In this manuscript, we made a first attempt at training samplers of the Edwards-Anderson model. This analysis will be continued in the future by investigating larger system sizes, higher values of inverse temperatures, and other types of couplings $J$. Since our method can be applied to any type of interactions of spins, other spin glass models can be considered. Another direction is related to improvements in the neural samplers that are based on some properties of the physical system. In this paper, we considered simple approximate probabilities, but this can be further explored. Another possibility is to combine transformers with methods proposed in the literature, such as HAN \cite{Bialas:2022qbs}, TwoBo \cite{Biazzo_2023,2024MLS&T...5b5074B}, or RiGCS \cite{Singha:2025lsd}. Finally, we should note that the transformer architecture applied in this work is very small compared to those used in modern LLMs. Therefore, natural questions arise: i) can larger spin systems be effectively sampled with larger architectures? ii) What kind of modifications of the  original proposal \cite{NIPS2017_3f5ee243} developed for LLMs in recent years can be applied to the neural samplers to improve their performance? iii) Can new improvements be developed that can be applied to this specific problem? To answer those questions, one should consider the differences between text or picture generation and the generation of spin systems with a given interaction. It's rather clear that correlations between generated tokens will differ for all these applications. For example, when a large spin system size is considered, the context of the attention mechanism should be very long. On the other hand, the correlation between distant tokens may be weak (at least when the system is not in the critical point), which suggests that the attention matrix may be sparse.    

It is clear that neural samplers are not currently competitive with the state-of-the-art simulation methods in spin systems (see, for example, review \cite{ALTIERI2024361} for modern methods to study spin glass systems). On the other hand, given the short history of these algorithms and the progress witnessed in recent years, we believe it is a promising direction for research.

\section*{Acknowledgments}
We thank Elia Cellini and Kim Nicoli for providing the details of the RiGCS algorithm training. We gratefully acknowledge Polish high-performance computing infrastructure PLGrid (HPC Center: ACK Cyfronet AGH) for providing computer facilities and support within computational grant no. PLG/2025/018811. We acknowledge Polish high-performance computing infrastructure PLGrid for awarding this project access to the LUMI supercomputer, owned by the EuroHPC Joint Undertaking, hosted by CSC (Finland) and the LUMI consortium through PLL/2025/08/018112. The study was funded by "Research support module" as part of the "Excellence Initiative -- Research University" program at the Jagiellonian University in Kraków.
T.S., A.S. and D.Z. kindly acknowledge the support of the Polish National Science Center (NCN) Grant No.\,2021/43/D/ST2/03375.  P.K. acknowledges the support of the Polish National Science Center (NCN) grant No. 2022/46/E/ST2/00346. 

\appendix

\section{Implementation details}
\label{impl_details}
The architecture of the transformer is a modified version of Ref. \cite{nanoGPT}, where the \texttt{PyTorch} library \cite{2019arXiv191201703P} was used. It is a decoder-only transformer with additive learnable positional encoding. We added the KV-cache to speed up generation time. The weights of all linear layers and embeddings were initialized with a normal distribution with an average $\mu = 0$ and a standard deviation $\sigma = 0.02$. Biased linear layers were used, with bias initialized to 0.  The detailed hyperparameters are shown in Tables \ref{tab:hyperparameters} and \ref{tab:hyperparameters_2}, where:
\begin{itemize}
    \item $n_\text{embed}$ is the size of the embedding space, 
    \item  $n_\text{layer}$ is the number of transformer layers, 
    \item  $n_\text{head}$ is the number of attention heads in each layer,
    \item "FFN" means feed-forward network.
\end{itemize}

\begin{table}[H]
    \begin{minipage}{.5\linewidth}
    \centering
    \begin{tabular}{l c}
        \toprule
        Hyperparameter & Value \\
        \midrule
        $n_\text{embed}$ & 128 \\
        $n_\text{layer}$ & 1 or 2  \\
        $n_\text{head}$ & 4  \\
        dropout & 0  \\
        FFN hidden dimension & $4 \times n_\text{embed}$  \\
        activation function & GeLU  \\
        \bottomrule
        
    \end{tabular}
    \caption{\centering{Architecture hyperparameters. See text for explanation of parameters. }}
    \label{tab:hyperparameters}
    \end{minipage}
    \begin{minipage}{.5\linewidth}
    \centering
    \begin{tabular}{l c}
        \toprule
        Hyperparameter & Value \\
        \midrule
        Optimizer & AdamW \\
        learning rate & 0.001, constant  \\
        $\beta_1$ & 0.9  \\
        $\beta_2$ & 0.95  \\
        batch size & 4096 - 8192 \\
        Hardware & $4 \times \text{AMD MI250x}$\\
        \bottomrule
        
    \end{tabular}
    \caption{Training hyperparameters.}
    \label{tab:hyperparameters_2}
    \end{minipage}
\end{table}
The mapping of tokens to patches was done in a little endian fashion before reshaping into rectangular region in a row-major order. Thus, the number is firstly changed into bits, with the least-significant being on the far left. Those bits are then mapped $\{0, 1\} \xrightarrow{} \{-1, 1\}$. After that, we reshape the constructed array so that consecutive numbers end up in the same row. Taking $2 \times 4$ patch as an example:

\begin{equation*}
\begin{split}
&0 \leftrightarrow{}
\begin{pmatrix}
-1 & -1 & -1 & -1\\
-1 & -1 & -1 & -1
\end{pmatrix}, \quad
 1 \leftrightarrow{}
\begin{pmatrix}
1 & -1 & -1 & -1\\
-1 & -1 & -1 & -1
\end{pmatrix},\quad
2 \leftrightarrow{} 
\begin{pmatrix}
-1 & 1 & -1 & -1\\
-1 & -1 & -1 & -1
\end{pmatrix}, \quad \\
&4 \leftrightarrow{}
\begin{pmatrix}
-1 & -1 & 1 & -1\\
-1 & -1 & -1 & -1
\end{pmatrix},\quad
31 \leftrightarrow{} 
\begin{pmatrix}
1 & 1 & 1 & 1\\
1 & -1 & -1 & -1
\end{pmatrix}.
\end{split}
\end{equation*}

\section{Scaling of tVAN's numerical cost}
\label{scaling_appendix}

\subsection{Number of operations for transformer}
\label{scaling_derivation_details}

Below, we derive the number of operations performed during one forward pass through the transformer, ignoring KV-cache. A similar derivation can be found in \cite{scaling-book}. We divide the transformer into basic components, and each of them is analyzed separately. We assume a batch size equal to one, as a larger batch size simply multiplies all factors by the proportional constant. The final result is a sum of all constituents.

We assume just one layer of the transformer, $n_\text{layer}=1$ (attention + feed-forward network).
We ignore contributions from element-wise operations, including activation functions. We track only the leading scaling of each component.

As in the main body of the draft, we denote: $N_{\text{context}}$ -- the length of the input sequence;
$N_\text{vocab}$ -- number of possible tokens.
The number of operations for each component of the transformer is:
\begin{enumerate}
    \item Embeddings -- this layer encodes tokens into $n_\text{embed}$-dimensional vectors, while also adding positional embedding.

    For every token, this requires a copy of one row of the $N_\text{vocab} \times n_\text{embed}$ matrix.

    Number of operations: $N_{\text{context}} \times n_{\text{embed}}$.

    \item Scaled dot-product attention -- $\text{softmax}\left(\frac{QK^T}{\sqrt{d_k}}\right)V$. We calculate as if there is just one head, but the number of operations stays the same even when we have multiple heads. This is because multi-head attention partitions the embedding dimension across heads, keeping the number of operations constant.
    \begin{itemize}
        \item Linear layer --
            it produces $K$, $Q$ and $V$ values. It is a matrix of size $\propto n_{\text{embed}} \times n_{\text{embed}}$.
            
        Number of operations: $n_{\text{embed}}^2 \times N_{\text{context}}$.

        \item Dot product -- $\sum_j Q_{ij} K_{kj}$, both Q, K are $N_{\text{context}} \times n_{\text{embed}}$ matrices.

        Number of operations:
        $N_{\text{context}}^2 \times n_{\text{embed}}$.
        \item Softmax -- calculated over the rows of the $N_{\text{context}} \times N_{\text{context}}$ matrix from the dot product.

    Number of operations: $N_{\text{context}}^2$.

    \item Matrix multiplication -- multiplication with V, matrices of sizes $N_{\text{context}} \times N_{\text{context}}$ and $N_{\text{context}} \times n_{\text{embed}}$.
    
    Number of operations:  $N_{\text{context}}^2\times n_{\text{embed}}$.

    \item Projection layer -- linear layer with a matrix of size $n_\text{embed} \times n_\text{embed}$.
    
    Number of operations: $n_{\text{embed}}^2 \times N_{\text{context}}$.
    \end{itemize}

    \item Feed-forward: two linear layers with a hidden dimension $\propto n_\text{embed}$.

    Number of operations: $n_{\text{embed}}^2 \times N_{\text{context}}$.
    \item Final linear layer with size $n_{\text{embed}} \times N_{\text{vocab}}$.
    
    Number of operations: $n_{\text{embed}} \times N_{\text{vocab}} \times N_{\text{context}}$.

    \item Final softmax.

    Number of operations: $N_{\text{context}} \times N_{\text{vocab}}$
\end{enumerate}

Altogether,
\begin{equation}
\label{eq-comp}
O\!\left(n_{\text{embed}} \times N_{\text{vocab}} \times N_{\text{context}}\right) + O\!\left(n_{\text{embed}}^2 \times N_{\text{context}} \right)+ O\!\left(N_{\text{context}}^2 \times n_{\text{embed}}\right).
\end{equation}

 The relative size of those three contributions is not easy to assess; hence, the precise determination of how the numerical cost of the algorithm depends on different hyperparameters is out of reach. Nevertheless, there are several conclusions we can draw from the above formula when applied to tVAN.

Substituting expressions for vocabulary size and context size, $N_{\rm vocab}=2^{r\times c}$ and  $N_{\rm context}=L^2/(r\times c)$ into Eq.~(\ref{eq-comp}) we get:

\begin{equation}
    O\left(n_{\rm embed} \times 2^{r\times c} \times \frac{L^2}{r\times c} \right)+ O\left(n_{\rm embed}^2 \times \frac{L^2}{r\times c} \right)+ 
    O\left(\frac{L^4}{(r\times c)^2} \times n_{\rm embed}\right).
    \label{scaling_formula}
\end{equation}

In the first term of this equation, the patch size $r\times c$ appears in the exponent (since $N_{\rm vocab}=2^{r\times c}$), whereas the other two terms have a power-like dependence on it.
As it is a competition between the exponential growth of vocabulary size and the quadratic (with $L$) growth of the number of generated tokens (context length), it is clear that patches need to be rather small compared to $L^2$. 
For large $L$, the third term dominates the second one, and comparing it with the first term, one sees the same scaling with the embedding dimension. Assuming that the optimal patch size is the one that leads to the same scaling with $L$ of both terms, we find the condition $2^{rc} \approx L^2/{rc}$. For $100\lesssim L \lesssim200$ this condition means $10\lesssim rc \lesssim 12$, which gives a rough estimate for the optimal patch size. This agrees with our findings after testing different patch sizes (see section \ref{prob_effect}).

\subsection{Numerical results}

We now present the scaling of the tVAN numerical cost with the system size $L$. We will focus on the physical time required to run one epoch of training and on the memory requirements of the algorithm. For simplicity, we do not take into account approximate probabilities, as they do not contribute much to the numerical cost of simulation, including them would not change the conclusions. For the comparison between different system sizes, we keep most of the hyperparameters the same, changing only the number of generated token sequences $N_{\rm context}$. The patch shape is fixed at $2\times 4$ and the batch size is 512.

\begin{figure}
\begin{subfigure}[b]{0.5\textwidth}
    \includegraphics[width=\textwidth]{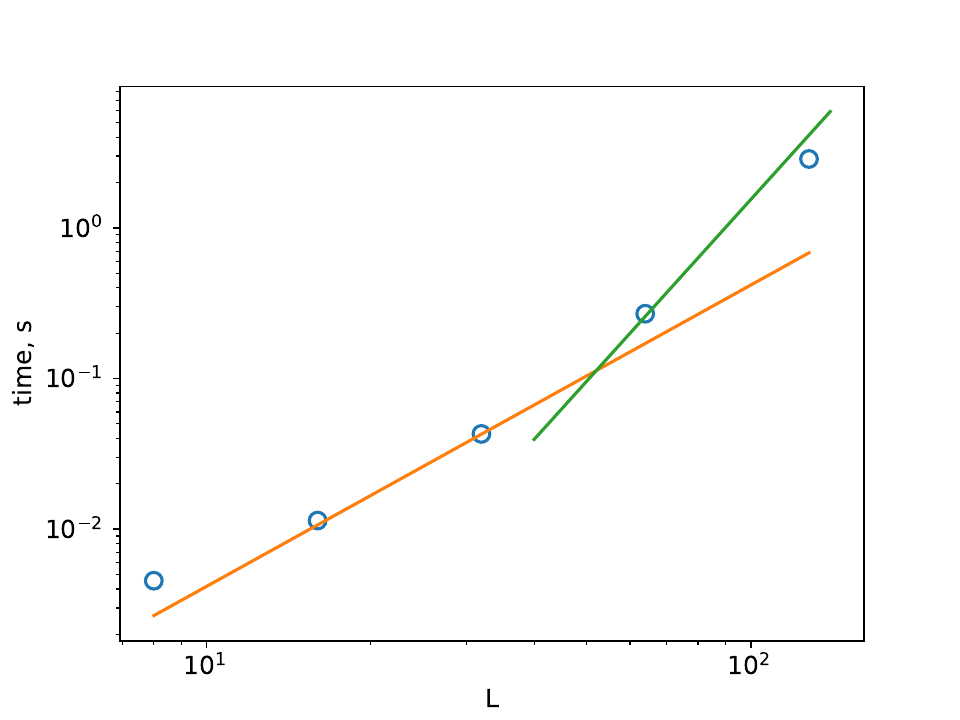}
\end{subfigure}
\begin{subfigure}[b]{0.5\textwidth}
    \includegraphics[width=\textwidth]{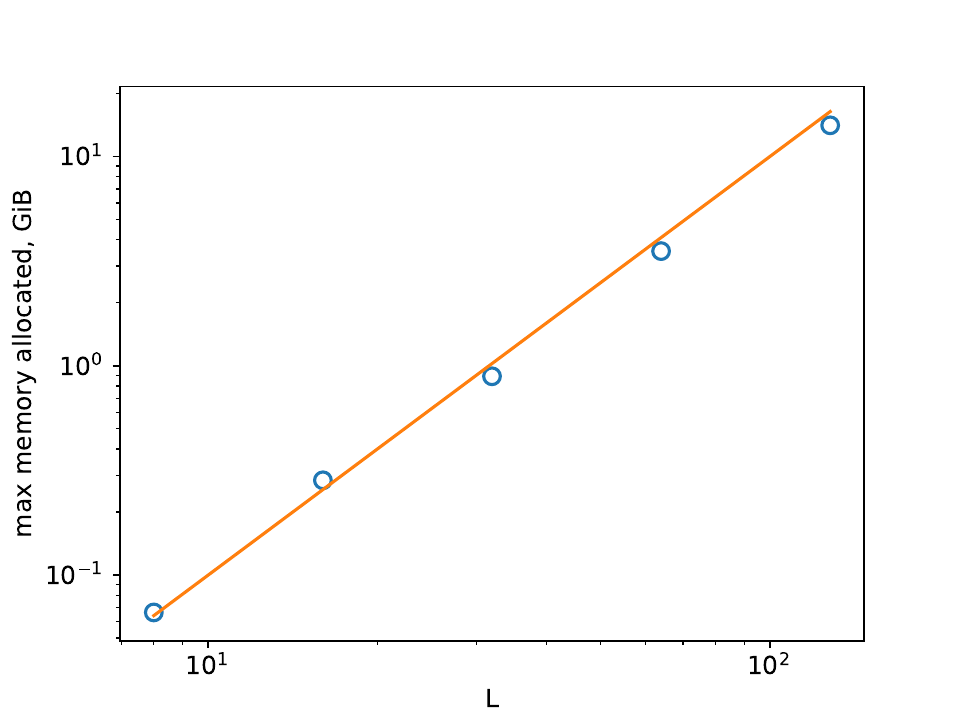}
\end{subfigure}
    \caption{
    Left: time of one epoch as function of system size $L$ (blue points). The lines shows behavior $\sim L^2$ (orange) and $\sim L^4$ (green). Right: Max memory allocated during transformer training depending on the system size with 2x4 patches and batch size of 512. Line shows $\sim L^2$ behavior.}
    \label{fig:time_memory}
\end{figure}

According to Eq.~\ref{scaling_formula}, the number of operations necessary to generate one configuration using the transformer should rise with $L$ at least as fast as $\sim L^2$ and not faster than $\sim L^4$.  We may expect that the time for one epoch of tVAN training will exhibit similar scaling if we assume that it is proportional to the number of operations needed to generate a configuration.
In the left panel of Fig.~\ref{fig:time_memory}, we show the wall time of one epoch as a function of $L$ (blue points). Since the scales on the axes are logarithmic, the power-like behavior would manifest as a line. For the convenience of the reader, we show the $\sim L^2$ and $\sim L^4$ dependence as lines (orange and green, respectively). In the range we considered, $8\le L\le 128$ we do not observe power-like behavior. At small $L$, the dependence is close to $\sim L^2$ whereas at large $L$, the rise is closer to $\sim L^4$ which agrees with expectations that $\sim L^4$ behavior should be dominant for large enough $L$. One can also see that the measured time for $L=8$ is somewhat larger than expected from $\sim L^2$ behavior - this is a sign of breaking of scaling \ref{scaling_formula}, which is due to overhead from other operations not taken into account in our simple estimation. For such small $L$ formally sub-leading terms $\sim L^0, \ \sim L^1$ may be important.

The second factor, which is crucial in neural network training, is memory allocation. The larger the model is, the less GPU memory is available for the spin configurations in the batch. Hence, for larger $L$, we are forced to use a smaller batch size, which reduces the efficiency of training. This problem is partially alleviated by data parallelism and the use of multiple GPU units. In the right panel of Fig.~\ref{fig:time_memory} we show maximal memory allocation (in GB) for the different $L$ values. We observe a rather stable $\sim L^2$ dependence over the whole range of  $8\le L\le 128$. 
Most of the memory is taken by the auxiliary variables used in calculations of gradients.  Those variables are created  dynamically during the forward pass through the transformer and predicting their exact size would be very hard. Hence, we 
present just the results of numerical experiments.

We note that the tests we performed here concern the time of one training epoch, but not the total training time. The latter is harder to measure, as it can vary even for the same architecture and hyperparameters, for example,  due to the different initialization of weights. But in general, the larger the system is, the more epochs are needed to train it. Hence, in practice, the total training time will scale faster with $L$ than $\sim L^2$ or $\sim L^4$. On the other hand, once trained, the algorithm can generate MC samples according to the scaling presented in this section. This will be the dominant cost of the whole simulation (training+ data generation) when reaching very large statistics is necessary.

\bibliography{references2.bib}

\end{document}